\documentclass[11pt]{iopart}

\usepackage[usenames,dvipsnames]{color}
\usepackage{graphicx}

\begin{document}

\title[Electronic structure of gapped PbPdO$_2$]{Hybrid functional electronic 
structure of PbPdO$_2$, a small-gap semiconductor}

\author{Joshua A. Kurzman, Mao-Sheng Miao, and Ram Seshadri}

\address{Department of Chemistry and Biochemistry, 
Materials Department and Materials Research Laboratory, 
University of California, Santa Barbara, CA 93106, USA}
\ead{jkurzman@chem.ucsb.edu}

\begin{abstract}
PbPdO$_2$, a ternary compound containing the lone-pair active ion Pb$^{2+}$
and the square-planar d$^8$ Pd$^{2+}$ ion, has attracted recent interest
because of the suggestion that its electronic structure, calculated within
density functional theory using either the local density or the generalized
gradient approximations, displays zero-gap behavior. In light of the potential
ease of doping magnetic ions in this structure, it has been suggested that 
the introduction of spin, in conjunction with zero band gap, can result in 
unusual magnetic ground states and unusual magnetotransport. It is known that 
most electronic structure calculations do not properly obtain a band gap even 
for the simple oxide PdO, and instead obtain a metal or a zero-gap 
semiconductor. Here we present density functional calculations employing 
a screened hybrid functional (HSE) which correctly obtain a band gap for 
the electronic structure of PdO. When employed to calculate the electronic 
ground state of PbPdO$_2$, a band gap is again obtained, which is consistent 
both with the experimental data on this compound, as well as a consideration 
of valence states and of metal-oxygen connectivity in the crystal structure. 
We also present comparisons of the absolute positions (relative to the vacuum 
level) of the conduction band minima and the valence band maxima in 
$\alpha$-PbO, PdO and PbPdO$_2$, which suggest ease of $p$-type doping in 
PbPdO$_2$ that has been observed even in nominally pure materials.

\end{abstract}


\maketitle

\section{Introduction}

Both lead and palladium -- in their own right -- present an array of 
fascinating phenomena in the oxide solid state, and find use in a host of 
important applications. The stereochemically active lone pair of Pb$^{2+}$ 
is associated with numerous electronic effects observed in complex oxides, 
such as the ferroelectric polarization of PbTiO$_3$, and the 
high piezoelectric response of PbZr$_{1-x}$Ti$_x$O$_3$ (PZT).\cite{Seshadri:2001}
Palladium is ubiquitously employed in both heterogeneous and homogeneous 
catalysis, and is a critical component in catalytic converters for automotive 
exhaust purification. The catalytic action of Pd$^{2+}$ ions in oxides has 
been a topic of much interest recently, addressed for example, by Hegde and 
co-workers.\cite{Hegde2009} Given the diversity of properties associated with 
Pb$^{2+}$ and Pd$^{2+}$ cations in oxides, it is of interest to examine how 
they behave together in a complex oxide.

\begin{figure}
\centering
\includegraphics[width=6in]{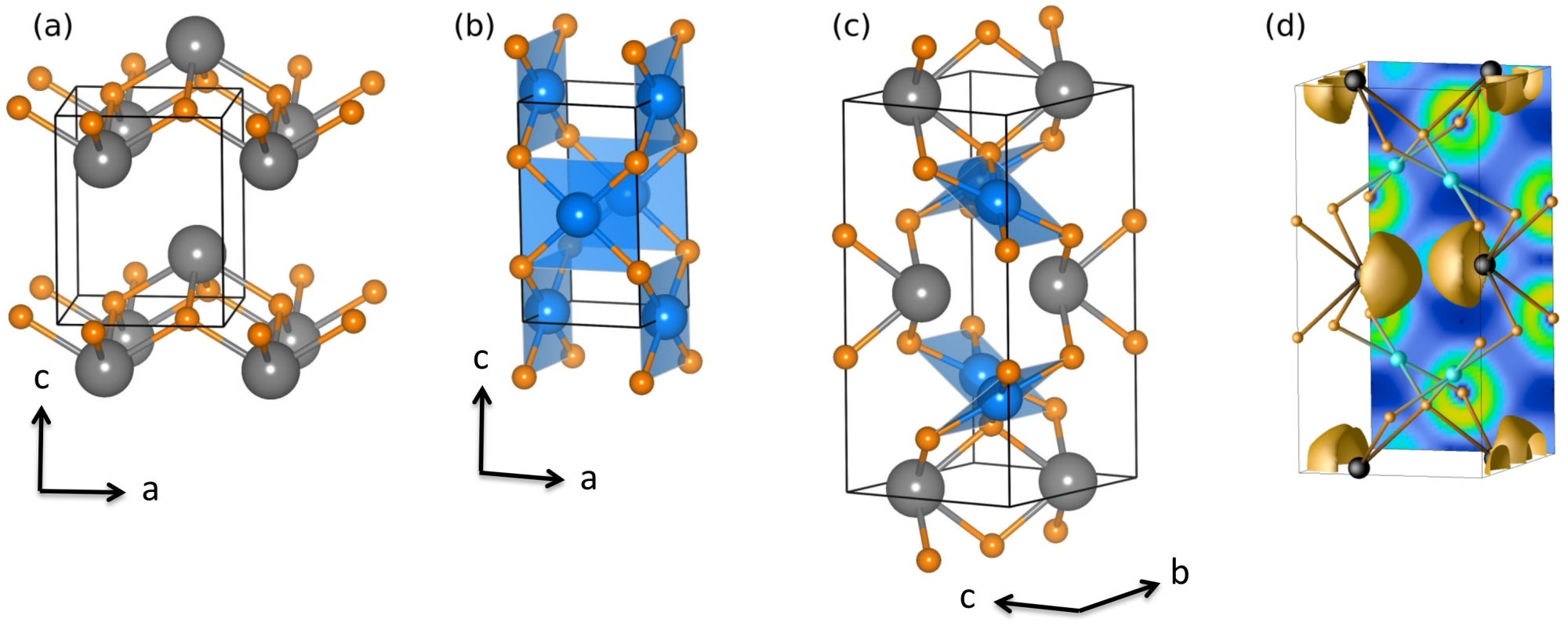}
\caption{(Colour online) Unit cell depictions, drawn to scale, of 
(a) $\alpha$-PbO ($P4/nmm$ \#129) (b) PdO ($P4_2/mmc$ \#131) and 
(c) PbPdO$_2$ ($Imma$ \#74). O atoms are orange, Pb atoms grey, and Pd atoms 
blue. PdO$_4$ square planes are shown in polyhedral rendering. (d) Electron 
localization function (ELF) isosurface projected within the unit cell of 
PbPdO$_2$ for an ELF value of 0.75.}
\label{fgr:structure}
\end{figure}

Litharge, the tetragonal form of PbO ($\alpha$-PbO, $P4/nmm$ \#129), is a red 
compound with a bandgap of about 1.9\,eV.\cite{Keezer:1968} Its crystal 
structure, depicted in Figure \ref{fgr:structure}(a), is layered, with
each layer comprising a square lattice of O that is alternately
(up and down) capped by Pb. The layers eclipse one-another when projected 
down the long ($c$) axis of the structure.  

Palladium oxide crystallizes in the tetragonal cooperite 
structure,\cite{Moore:1941} (space group $P4_2/mmc$ \#131), and consists of 
edge-sharing chains of PdO$_4$ planes that run perpendicular along the 
$a$ axis and the $b$ axis of the structure, and these chains corner-share
along the $c$ axis [Figure \ref{fgr:structure}(b)]. All of the Pd--O bonds 
are equidistant, but the PdO$_4$ units are not perfect squares and instead
are rectangular, with a slight compression leading to shorter nearest-neighbor 
O--O distances along $c$ than in the $a$ direction. Oxygen is tetrahedrally 
coordinated by four Pd atoms in the structure. The nearest-neighbor Pd--Pd 
distance of slightly over 3\,\AA\/ (the $a$ axis length) is relatively short. 
These short contacts occur in two dimensions in PdO, within and between the 
chains, and should give rise to disperse bands derived from the Pd d orbitals.

Rogers \textit{et al.}\ reported that single crystals of PdO are dark 
green,\cite{Rogers:1971} but the presence of impurities prevented the reliable 
determination of the band gap. PdO displays $p$-type semi-conductivity, and 
it has been suggested that the hole population arises from extrinsic defects, 
in particular the presence of unintentional dopants.\cite{Rogers:1971} Indeed, 
widely varying estimates of the PdO band gap have been reported: Hulliger 
found an optical band gap of 0.6\,eV from diffuse-reflectance measurements on 
a polycrystalline powder.\cite{Hulliger:1965} Okamoto and Aso estimated the 
band gap to be 1.5\,eV from conductance measurements of 
PdO films,\cite{Okamoto:1967} in agreement with the gap reported by 
Pillo \textit{et al}.\ from photoemission studies.\cite{Pillo:1997} 
Nilsson and Shivaraman found a 0.8\,eV gap from optical transmittance 
measurements of PdO films,\cite{Nilsson:1979} while a study of PdO films by 
Rey \textit{et al}.\  yielded an optical gap of about 2.1\,eV.\cite{REY:1978} 
 
PbPdO$_2$ is the only ternary oxide of Pb and Pd reported in the Inorganic 
Crystal Structures Database.\cite{Belsky:2002} Single crystals of PbPdO$_2$ 
were prepared by Meyer and M\"uller-Buschbaum, which they observed to be 
black.\cite{MEYER:1978} \textit{Locally}, the Pb-capped planes in $\alpha$-PbO and roughly square planar coordination of Pd in PdO are preserved in orthorhombic PbPdO$_2$, which was determined to 
crystallize in the $Imma$ \#74 space group. However, the relationship between 
the structural units is very distinct in this ternary phase 
[Figure \ref{fgr:structure}(c)]. PbPdO$_2$ can be viewed as a layered 
structure containing zig-zagging sheets of corner-sharing PdO$_4$ planes, 
interleaved by layers of Pb. Oxygen is tetrahedrally coordinated to two 
Pd within the sheets and two Pd between the sheets. Between the Pd layers, 
edge-sharing PbO$_4$ units form 1D chains running parallel to $b$ with Pb 
on alternating sides of the chain, and the chains are eclipsed with respect 
to one another along the $c$ axis.

While an understanding of the PbPdO$_2$ structure is facilitated by describing 
it as a layered material, it is important to point out that chemically it 
should be regarded as a three-dimensional extended solid.  The Pd--O and Pb--O bonds in the simple oxides of Pd and Pb are quite covalent, and PbPdO$_2$ contains an extended network of 
--Pd--O--Pb--O--Pd--  connectivities. There are two points that we emphasize 
here about the structure of PbPdO$_2$: First, the dimensionality of the 
PdO$_4$--PdO$_4$ and PbO$_4$--PbO$_4$ connectivities are reduced in PbPdO$_2$, 
relative to those in the oxides PdO and $\alpha$-PbO. Second, there 
are no edge-sharing PdO$_4$ units present in PbPdO$_2$ or any short 
Pd--Pd distances, as encountered in PdO, nor are there any in-plane 
Pd--O--Pd connectivities. Qualitatively, considering the structural 
arrangement and electronic configurations of the cations, one would expect 
PbPdO$_2$ to not be any closer to metallic behavior than the two 
binary compounds that it is derived from.

Our interest in PbPdO$_2$ stems from prior and ongoing efforts to understand 
relationships between the electronic structural features of platinum group and 
noble metal ions, and the catalytic activity of the oxides they form. 
We recently reported an experimental and computational comparison of the 
complex oxides of isoelectronic Pd$^{2+}$ and Au$^{3+}$, 
La$_2$BaPdO$_5$ and La$_4$LiAuO$_8$, both of which contain isolated PdO$_4$ 
and AuO$_4$ square planes, respectively.\cite{Kurzman:2010fk} Density 
functional calculations within the local density approximation (LDA) 
yielded sizeable (though underestimated) band gaps for both compounds, 
consistent with their yellow colors and optical absorption spectra. That 
LDA was a sufficient level at which to study these compounds is related to 
the isolated nature of the Pd and Au coordination environments. In contrast,
although PdO is a well known semiconductor, LDA calculations often predict 
it to be metallic.\cite{PARK:1994}

The electronic structure of PdO has been previously studied by a variety of
\textit{ab-initio} methods. Hass \textit{et al}.\ reported the first
computational investigation of PdO, in which an augmented-spherical-wave
approach within the local density approximation was
employed.\cite{HASS:1992} PdO was correctly predicted to be a semiconductor,
with a direct gap at the $M$ point, although the size of the gap was
severely underestimated. A subsequent study reported by Park \textit{et
al.}, using full-potential approaches with linearized augmented-plane-waves
(LAPW) and linear-muffin-tin-orbitals (LMTO) within the LDA, was unable to
predict a gap in PdO,\cite{PARK:1994} though a small gap was reported to
open through the use of LMTO in the
atomic-sphere-approximation.\cite{AHUJA:1994}  Uddin \textit{et al}.\
reported a study of density functional predictions for the
electronic structure of PtO,\cite{Uddin:2005} which is isostructural with
PdO. Using Gaussian-type orbitals, the Heyd, Scuseria, and Ernzerhof
screened hybrid functional\cite{Heyd:2003} correctly predicted PtO to be a
semiconductor, whereas calculations within the local spin-density
approximation (LSDA), Perdew Burke and Ernzerhof (PBE) generalized gradient
approximation (GGA), and with a meta-GGA functional, all led to the
prediction of (poor) metallicity.  Recently, a HSE band gap of 0.9\,eV in 
PdO has been reported by Seriani \textit{et al.},\cite{Seriani:2009} and 
Bruska \textit{et al}. determined an HSE band gap of 0.8\,eV that could
be extended to 1.5\,eV by increasing the amount of Fock exact exchange 
to 39\,\%.\cite{Bruska:2011}

On the basis of LDA calculations, Wang suggested that PbPdO$_2$ is a
zero-gap oxide semiconductor,\cite{Wang:2008} with potentially very interesting
electronic structural signatures when the compound is subject to partial
substitution by magnetic ions. Chen \textit{et al.}\ reported a combined 
experimental and computational study of PbPdO$_2$, employing the PBE 
functional with the addition of a 2\,eV Hubbard $U$ term on Pd to account 
for on-site Coulomb repulsion.\cite{Chen:2011} Their GGA+$U$ calculations 
did not result in the opening of a band gap. However, 
Bennett \textit{et al}.\ recently noted that an exceedingly large $U$ value of 
8\,eV, using the LDA+$U$ method, was needed to reproduce the experimental 
gap of PdO.\cite{Bennett:2010}  

Electrical transport properties of polycrystalline PbPdO$_2$ were reported
by Ozawa \textit{et al.},\cite{Ozawa:2005} and Seebeck and Hall
effect measurements indicated $p$-type conductivity. It was suggested 
from a change in the temperature coefficient of resistance that PbPdO$_2$ 
undergoes a metal-to-insulator transition at about 90\,K, and that the 
ground state is an insulator, not a metal. However, the minimum resistivity 
of a little over 0.4\,$\Omega$cm is more consistent with semiconducting 
behavior than it is with the resistivity of a metal, which is expected to 
be of the order of 10$^{-2}$\,$\Omega$cm or less. Additionally, the reported 
change in resistivity between the ``insulating'' and ``metallic'' states is 
less than a factor of 2, whereas typical metal-to-insulator transitions 
span two or more orders of magnitude. For comparison, Uriu \textit{et al.}\ 
studied the effect of Li-substitution in PdO (Pd$_{1-x}$Li$_x$O$_{1-\delta}$), 
and found that the resistivity at 273\,K decreased from 3\,$\Omega$cm 
in the unsubstituted material, to $2.2\times 10^{-2}$\,$\Omega$cm with 
1\,mol\% Li, to $4.4\times 10^{-3}$\,$\Omega$cm with 
8\,mol\% Li.\cite{URIU:1991}

Reports of chemical substitution in PbPdO$_2$ -- with the intent of introducing electrical carriers and/or magnetic ions -- are few. Ozawa \textit{et al}.\ studied the substitution of Cu for Pd, which increases the resistivity.\cite{Ozawa:2005a}  Magnetism and transport have been examined in Co and Mn substituted PbPdO$_2$ by Lee \textit{et al}.\cite{Lee:2010,Lee:2011} 
Both Co and Mn lower the resistivity, but the minima are pushed to higher and lower temperatures, respectively, relative to unsubstituted PbPdO$_2$. The resistivity of PbPd$_{0.9}$Co$_{0.1}$O$_2$ is larger than 0.1\,$\Omega$cm at 150\,K, while for PbPd$_{0.9}$Mn$_{0.1}$O$_2$ the resistivity is 0.08\,$\Omega$cm at 70\,K.\cite{Lee:2011}  Large electroresistance and magnetoresistance in 
PbPd$_{0.75}$Co$_{0.25}$O$_2$ films have been reported by Wang \textit{et al.},\cite{Wang:2009} and quite recently, magnetism and transport properties of a PbPd$_{0.81}$Co$_{0.19}$O$_2$ film were studied by Su \textit{et al.}\cite{Su:2011}

In the light of the electronic properties that would be expected for
PbPdO$_2$ on the basis of its structural attributes and electron count, we
report density functional electronic structure calculations for PbPdO$_2$
employing a screened hybrid functional (HSE06). Inclusion of exact exchange
to the description of the short-range Coulomb potential has been shown to
significantly enhance the accuracy of density functional calculations in
periodic systems.\cite{Heyd:2004,Heyd:2005,Janesko:2009,Henderson:2011} To complement these results, 
we also present the HSE electronic structure of PdO. Additionally, by 
comparing the absolute positions, relative to the vacuum level, of the 
conduction band minima and the valence band maxima in $\alpha$-PbO, PdO and 
PbPdO$_2$, we find that facile electron transfer from PbPdO$_2$ to, for 
example, residual PdO left from the preparation, may account for the $p$-type 
conductivity observed in nominally pure materials. The HSE calculations
suggest surprisingly linear dispersions of the valence and conduction bands
across the direct $\approx$1.1\,eV band gap of this compound.

\section{Computational methods}

Density functional electronic structure calculations were performed in the
Vienna \textit{ab initio} Simulation Package
(VASP).\cite{Kresse:1994fk,Kresse:1996uq} Interaction between the valence
and core electrons was described using the projector augmented wave (PAW)
approach,\cite{Blochl:1994vn,Kresse:1999} with valence electrons described
by plane-wave basis sets. 14 valence electrons were included for Pb
(5d$^{10}$6s$^{2}$6p$^2$), 10 valence electrons for Pd (4d$^{10}$), and 6
valence electrons for O (2s$^2$p$^4$). Calculations were performed at the
scalar relativistic level, without the inclusion of spin-orbit coupling,
using both the GGA of Perdew, Burke, and Ernzerhof
(PBE),\cite{Perdew:1996kx} and the hybrid functional described by Heyd,
Scuseria, and Ernzerhof (HSE06).\cite{Heyd:2003,Krakau:2006}  The screened
hybrid functional introduces a percentage of exact nonlocal Fock exchange
(25\,\%) to the PBE functional, and a screening of 0.2\,\AA\/$^{-1}$ is
applied to partition the Coulomb potential into long-range and short-range
terms. Hartree-Fock and PBE exchange mix in the short-range portion, with
the long-range exchange represented by the PBE functional.

A cutoff
energy of 520\,eV was used in all calculations, as were $\Gamma$-centered $k$-point grids. Relaxations were deemed to have converged when the forces on all the atoms were less than 0.01\,eV\,\AA\/$^{-1}$.   Geometry optimization of PbPdO$_2$ was performed in the primitive
rhombohedral unit cell, using a $5\times5\times5$ $k$-mesh. The same grid was also used to calculate the HSE electronic density of states (DOS).  PBE calculations of the PbPdO$_2$ DOS employed a $20\times20\times20$ $k$-mesh. Band structure calculations for PbPdO$_2$ were performed using the conventional orthorhombic unit cell.
  For PdO, a
$6\times6\times4$ $k$-mesh was used to relax the structure with the HSE
functional and to calculate the DOS.  PBE relaxation and DOS calculations for PdO employed $8\times8\times6$ and $18\times18\times16$ $k$-point meshes, respectively.
PBE band structures were calculated with
30-point interpolations between each high symmetry $k$ point.  HSE band
structure calculations are presented only along the dispersions in the
region of the band gaps, $Z$ to $\Gamma$ to $X$ for PbPdO$_2$, and $X$ to
$M$ to $\Gamma$ for PdO, using 10-point interpolations between each high
symmetry $k$-point.

To investigate the doping tendencies of PbPdO$_2$, the positions of the
valence band maximum (VBM) and the conduction band minimum (CBM), relative
to vacuum, were studied. Slab models were constructed and atomic positions
optimized using the PBE functional with fixed lattice parameters. Four
Pb--O--Pd--O bilayers were included in PbPdO$_2$ slab, eight Pd--O layers
were used in the PdO slab, and eight PbO layers were used for the PbO slab.
The lengths of the ideal slabs for the three compounds are 19.20\,\AA\/,
21.82\,\AA\/ and 17.80\,\AA\/, respectively. Vacuum regions with the same
length as the slab were included for each compound.  Because the local and
semi-local PBE functional usually generates band gaps significantly smaller
than the experimental values, the hybrid HSE functional was employed for
static calculations of the PBE relaxed slabs.  To obtain the positions of
the VBM and CBM relative to the vacuum level, the electrostatic potential
in the center of the slabs (assuming it well represents the bulk) were
aligned with the center of the vacuum region.

Linear Muffin-Tin Orbital (LMTO)
calculation,\cite{Andersen:1975,Jepsen:1995} performed within the atomic
sphere approximation using version 47C of the Stuttgart TB-LMTO-ASA
program,\cite{Jepsen:2000} was used in the determination of the electron
localization function (ELF) for PbPdO$_2$.  Scalar-relativistic Kohn--Sham
equations within the local density approximation \cite{Barth:1972} were
solved taking all relativistic effects into account except for the
spin-orbit coupling.  The calculation was performed on 527 irreducible
$k$-points within the conventional orthorhombic cell. The ELF provides a
measure of the local influence of the Pauli repulsion on the behavior of
electrons, enabling the real-space mapping of core, bonding, and
non-bonding regions in a crystal.\cite{Becke:1990,Silvi:1994}

\section{Results and disucssion} 

\begin{figure}
\centering
\includegraphics[width=6in]{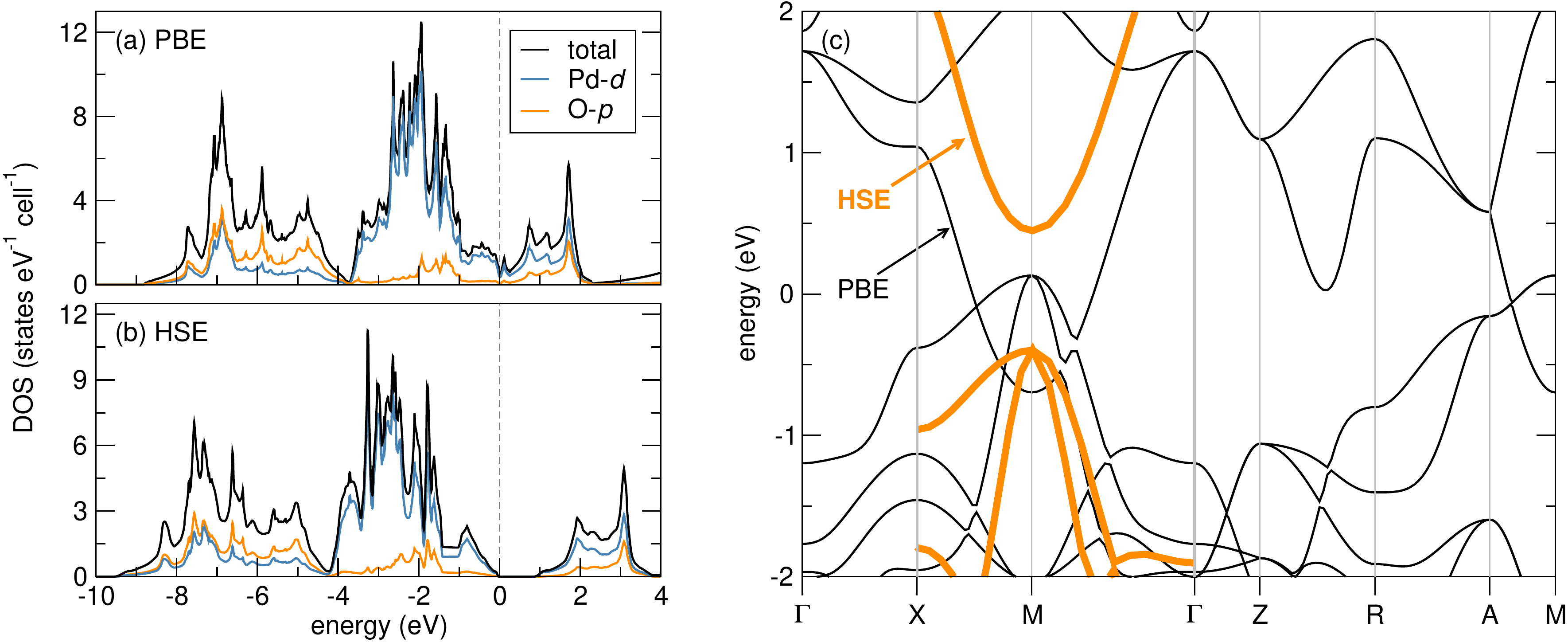}
\caption{(Colour online) PdO electronic density of states, orbital
projected density of states, and band structure. (a) PBE total and
projected DOS. (b) HSE total and projected DOS. (c) PBE band structure,
with HSE band structure overlaid in bold (orange) lines along the $X$ to
$M$ to $\Gamma$ dispersion. The direct band gap occurs at the $M$ point. In
(a) and (b) the Fermi energy is referenced to the top of the valence band,
indicated with vertical dashed lines. In (c), 0\,eV represents the Fermi
energy for the PBE calculation, while for the HSE bands the Fermi energy
has been adjusted to the absolute energy of the PBE $E_F$.   As has been
similarly noted by Uddin \textit{et al.}\ for PtO,\cite{Uddin:2005} the GGA
calculation does not predict PdO to be a semiconductor.}
\label{fgr:PdO}
\end{figure}

The total electronic and Pd-d and O-p orbital projected density of states
for PdO, calculated using the PBE and HSE functionals, are compared in
panels (a) and (b) of Figure \ref{fgr:PdO}. The PBE band-structure is shown
in panel (c) in the region of the valence band maximum (VBM) and conduction
band minimum (CBM), with bands from the HSE calculation overlaid in bold
lines along the dispersion from $X$ to $M$ to $\Gamma$. Structural parameters
obtained for the PBE and HSE relaxed structures are listed in Table
\,\ref{tab:params}, along with experimentally determined values for
comparison. Oxygen p-states lie predominantly between $-$9\,eV and
$-$4\,eV, with Pd d-states more densely populated between $-$4\,eV and the
Fermi energy, referenced to the top of the valence band in the DOS plots
and denoted with dashed vertical lines. In the plot of the band structure,
0\,eV represents the Fermi energy for the PBE calculation. The Fermi energy
of bands calculated with HSE have been adjusted to the absolute energy of
the PBE $E_F$.  Stabilization of the VBM and destabilization of the CBM by
HSE occur in almost equal amounts. Note the wide energy dispersion of bands
at the top of the valence band and bottom of conduction band, derived from
Pd d$_{xy}$ and d$_{x^2-y^2}$ orbitals, respectively.

The use of PBE leads to the prediction of poorly metallic character in
PdO.  HSE, on the other hand, correctly predicts PdO to be a semiconductor.
What we emphasize here is not the magnitude of the gap predicted by HSE,
which is a little less than 1\,eV, but rather that HSE affords a
qualitative prediction that agrees well with the known electronic structure
of PdO. The size of the gap is of the right order, but given the
disparities in the experimental gaps reported for PdO we cannot comment on
whether the prediction is an over- or under-estimate.

\begin{table} 
\caption{\label{tab:params}Structural parameters for PBE and HSE relaxed
PdO and PbPdO$_2$, compared with experimental values.  The experimental
parameters given are those determined by Rogers \textit{et al}.\ for
PdO,\cite{Rogers:1971} and Ozawa \textit{et al}.\ for
PbPdO$_2$.\cite{Ozawa:2005}  PbPdO$_2$, space group $Imma$ (\#129, setting
1): Pb (4$e$) at $0,1/4,z$ ; Pd (4$c$) at $1/4,1/4,1/4$ ; O (8$f$) at
$x,0,0$.  PdO, space group $P4_2/mmc$ (\#131): Pd (2$d$) at $1/2,0,0$ ; O
(2$e$) at $0,0,1/4$.}  \begin{indented} 
\item[]\begin{tabular}{@{}llll || lll} 
\br
 & PBE & HSE & expt. &  PBE & HSE & expt. \\
\mr
  & PdO & &  &   PbPdO$_2$ & & \\
\mr
$a$ (\AA\/) & 3.10   & 3.07  & 3.043  & 9.61 & 9.46 & 9.455 \\
 $b$ (\AA\/) &&& & 5.53 & 5.45 & 5.460 \\ 
 $c$ (\AA\/) &  5.44 & 5.33  & 5.336 & 4.77  & 4.77  & 4.660  \\
 Pb $z$ &&& & 0.772  & 0.764  & 0.774  \\
 O $x$ &&& & 0.350  & 0.350  & 0.345  \\
 Pd--O (\AA\/)& 2.06  & 2.03  & 2.02 & 2.06  & 2.04  & 2.01  \\
 Pb--O (\AA\/) &&& &  2.38 & 2.34  & 2.37  \\
 O--Pd--O $\angle$ ($^{\circ}$)& 97.5 & 98.1 & 97.5 & 95.8 & 96.2 & 94.3  \\
 \br 
 \end{tabular} 
 \end{indented} 
 \end{table}

\begin{figure}
\centering
\includegraphics[width=6in]{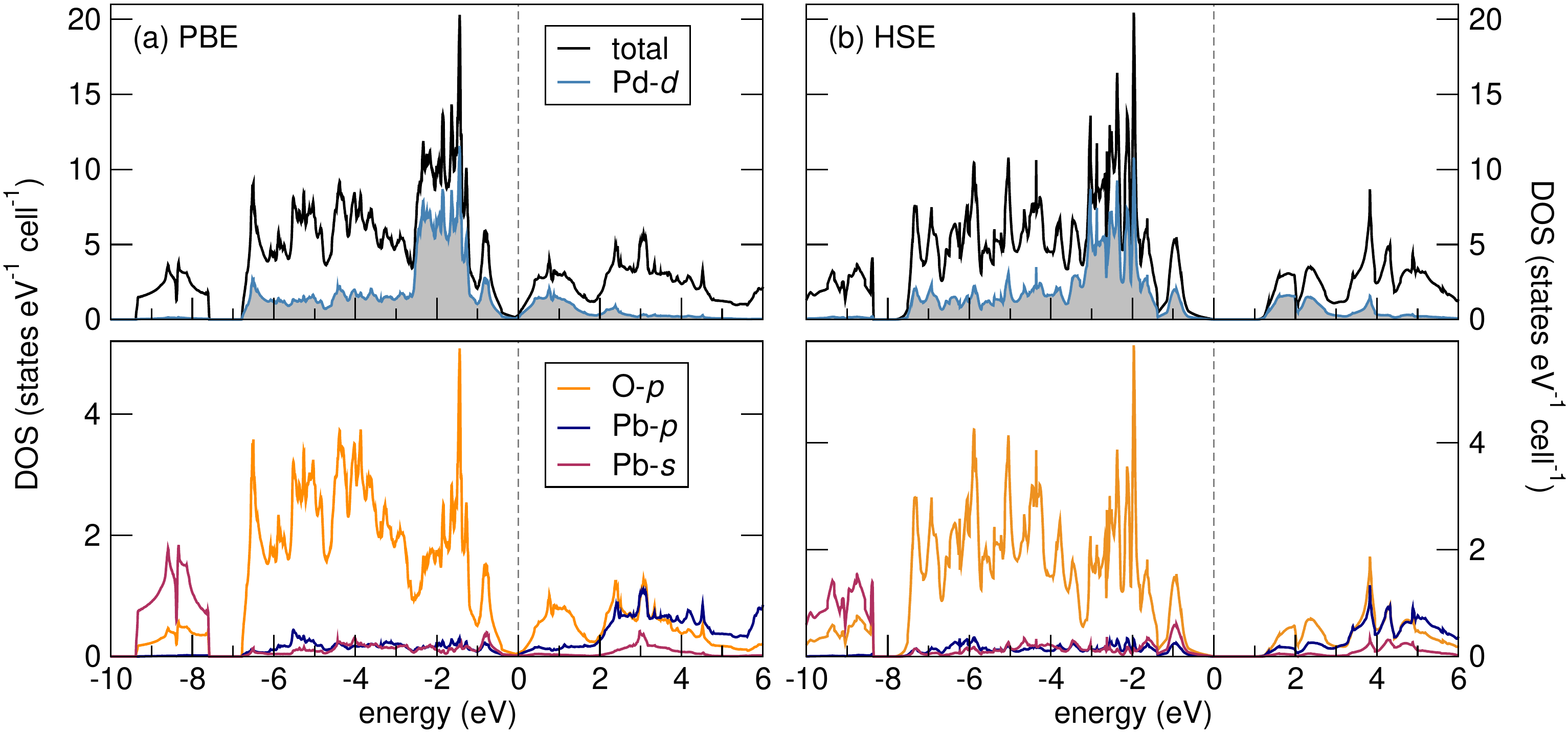}
\caption{(Colour online) Electronic density of states and orbital projected
density of states for PbPdO$_2$.  (a) PBE total and projected DOS. (b) HSE
total and projected DOS. The Fermi energy is referenced to the top of the
valence band, denoted by dashed vertical lines. The upper panels display the
total and Pd-d projected DOS, which are shaded for clarity, and the lower
panels show the O-p, Pb-p and Pb-s projected DOS.  The band gap predicted by
HSE is about 1.1\,eV. Note the strong hybridization of O-p and Pb-s states
between $-$10\,eV and $-$8\,eV, and the presence of Pb-s and p states near
the Fermi energy.}
\label{fgr:dos}
\end{figure}

Geometry-optimized structural parameters obtained with the PBE and HSE
functionals are given in Table \ref{tab:params}, and the total electronic
and orbital projected DOS of PbPdO$_2$ are presented in Figure
\ref{fgr:dos}. The shape and dispersion of the orbital contributions
obtained by the different methods are quite similar, with the obvious
difference being the presence of a $\approx$1.1\,eV bandgap in the HSE
electronic structure. The upper panels display the total and Pd-d projected
DOS, which are shaded for clarity, and the lower panels show the O-p, Pb-p
and Pb-s projected DOS.  The bottom of the Pd-d band in PbPdO$_2$ is
somewhat destabilized (closer in energy to $E_F$), relative to the
dispersion of the d states in PdO, and this is consistent with decreased
overlap associated with the absence of short Pd--Pd distances.  Note the
strong hybridization of O-p and Pb-s states between $-$10\,eV and $-$8\,eV,
and overlap of O-p and Pb-s and p states states near the Fermi energy,
features that are similarly encountered in the electronic structure of
$\alpha$-PbO.\cite{Watson:1999,Watson:1999a}  In $\alpha$-PbO, the highly
stabilized Pb-s and O-p states held deep below the $E_F$ are strongly
bonding, while the states just below $E_F$ are of antibonding
character.\cite{Raulot:2002}  The mixing of Pb-s and O-p states near the
Fermi energy has been suggested to drive the localization of the Pb lone
pair into lobe-shaped structures.\cite{Watson:1999,Watson:1999a}  As can be
seen in the electron localization function (ELF) isosurface of PbPdO$_2$
shown in panel (d) of Figure \ref{fgr:structure}, the Pb lone pair lobes
are localized apical to the base of the Pb-capped planes.

\begin{figure}
\centering
\includegraphics[height=3in]{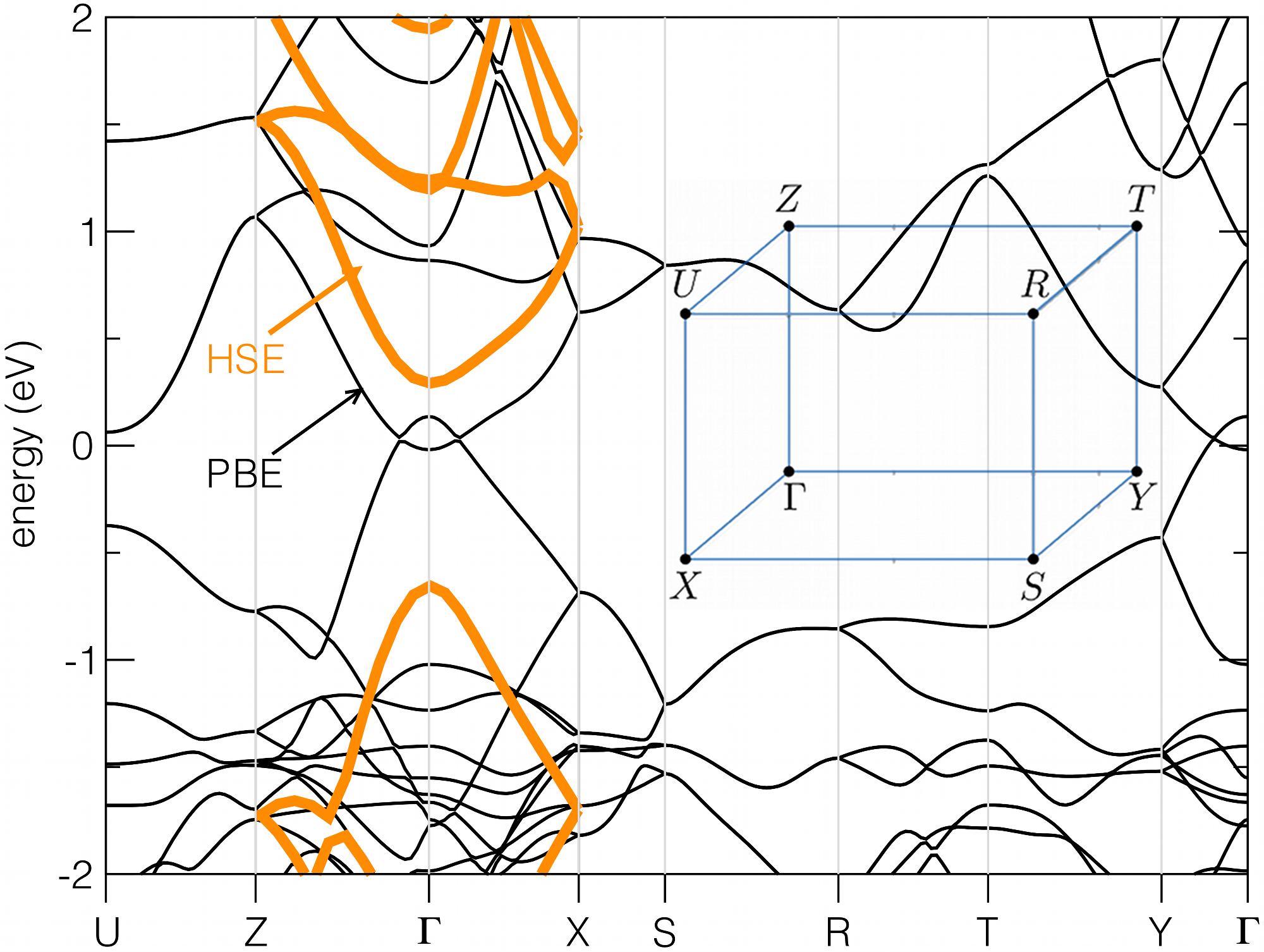} 
\caption{(Colour online) PBE calculated band structure of PbPdO$_2$ with the
HSE band structure overlaid in bold (orange) lines along the $Z$ to $\Gamma$
to $X$ dispersion.  The direct band gap occurs at the $\Gamma$ point.  0\,eV
references the Fermi energy obtained in the PBE calculation; the HSE bands
have been scaled to the same absolute reference energy. Note that the HSE
correction stabilizes (pushes down in energy) the top of the valence band.
The Brillioun zone of a primitive orthorhombic cell is inset in the figure.}
\label{fgr:bnds}
\end{figure}

The PBE band structure of PbPdO$_2$ in the vicinity of the Fermi energy is
shown in Figure \ref{fgr:bnds}, with the HSE calculated bands overlaid in
bold along the $Z$ to $\Gamma$ to $X$ dispersion. For the PBE calculation
0\,eV is taken as the Fermi energy, while for the HSE calculation the Fermi
energy has been adjusted to the absolute energy of the PBE $E_F$. Band
structure calculations were performed using the Brillioun zone (BZ) for a
conventional orthorhombic cell, and the BZ is inset in Figure \ref{fgr:bnds} for convenience.  
The PBE
calculated band structure shows a slight crossing of bands at the $\Gamma$
point, where a zero-gap was predicted by Wang.\cite{Wang:2008}  A direct
gap is clearly evident in the dispersion calculated with HSE. It is
important to note that the bands at the top of the valence band and at the
bottom of the conduction band do not show nearly as much energy dispersion
as the equivalent bands in PdO. Also noted is the surprising linearity of
the valence and conduction band dispersions near the direct band gap. 

\begin{figure}
\centering
\includegraphics[width=3in]{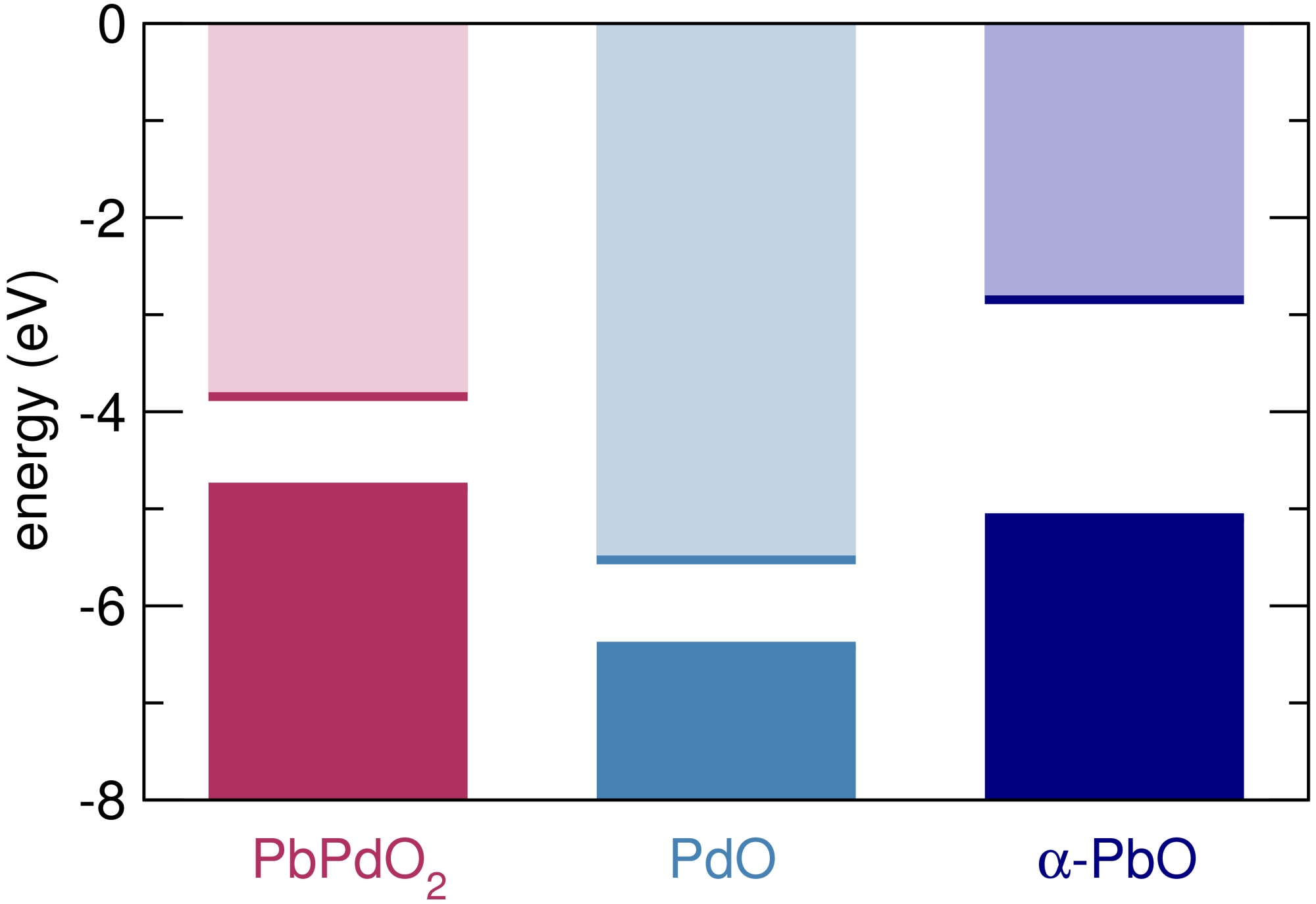} 
\caption{Schematic representation of the positions of valence band maxima
(VBM) and conduction band minima (CBM) in PbPdO$_2$, PdO, and $\alpha$PbO,
referenced to the vacuum level. Because the CBM of PdO lies below the VBM
of PbPdO$_2$, PbPdO$_2$ will be readily hole-dopeable in the presence of
even small amounts of residual PdO.}
\label{fgr:slabs}
\end{figure}

In Figure \ref{fgr:slabs} we present a schematic showing the calculated
positions of the valence band maxima (VBM) and conduction band minima (CBM)
for PbPdO$_2$, PdO, and $\alpha$-PbO. The VBM and CBM positions of
PbPdO$_2$ are 4.78\,eV and 3.84\,eV below vacuum, respectively, which are
quite high. This indicates that the compound is prone to $p$-type doping by
native defects and intentional substituents. Conversely, the VBM and CBM of
PdO are 6.42\,eV and 5.53\,eV below the vacuum, indicating that PdO is, in
comparison with PbPdO$_2$, more prone to being $n$-doped. More importantly,
the CBM of PdO is lower in energy than the VBM of PbPdO$_2$. Thus, the
presence of PdO in a sample of PbPdO$_2$ can lead to hole-doping of
PbPdO$_2$ by facile electron transfer from its valence band to the conduction band
of PdO.  Even small amounts of PdO would lead to the generation of a large
quantity of holes in the valence band of PbPdO$_2$, which may explain the
$p$-type conductivity observed for samples of PbPdO$_2$.

On the other hand, the VBM and CBM for $\alpha$-PbO are located at 5.09\,eV and 2.85\,eV
below vacuum, respectively. Because PbO is more ionic, it has a larger
gap, and its VBM is lower than the VBM of PbPdO$_2$ while the CBM of PbO is
higher than the CBM of PbPdO$_2$. Therefore, the presence of unreacted PbO
in samples of PbPdO$_2$ are not likely to significantly affect its
electronic properties, such as the density of charge carriers. Given the
volatility of PbO, PdO is much more likely to persist in samples of
PbPdO$_2$.

Our calculations only address alignment of the energy levels of the
starting materials typically used in the preparation of PbPdO$_2$, relative
to the target phase. A complete discussion of the source of hole
conductivity requires a comprehensive study of the defect and impurity
properties of the system -- as they may contribute greatly to the source
of the $p$-type doping -- which is beyond the scope of the present work.
However, our results reveal that PbPdO$_2$ is susceptible to $p$-type
doping, and emphasizes the importance of preparing high purity materials as
small amounts of residual starting material, PdO in particular, may greatly
influence the physical properties observed in this system.

\section{Conclusions}

In conclusion, electronic structure calculations carried out on PbPdO$_2$
at the hybrid HSE level suggest that the ground state of PbPdO$_2$
is actually that of a small band gap semiconductor. Slab calculations
to obtain the locations of the valence and conduction bands with respect to the
vacuum level suggest that PbPdO$_2$ can be easily hole-doped. Both these
observations are consistent with experimental observations on the system.
The valence band dispersion in this easily hole-doped material is surprisingly 
linear, potentially suggesting interesting low-temperature physics in this compound.

\section*{Acknowledgments}
We thank the Department of Energy, Office of Basic Energy Sciences for
support of this work through Grant No.\ DE-FG02-10ER16081. This work made
use of the computing facilities of the Center for Scientific Computing
supported by the California Nanosystems Institute, Hewlett Packard, and by
the Materials Research Laboratory at UCSB: an NSF MRSEC (DMR-1121053). JAK
thanks Kris Delaney and David Scanlon for helpful discussions pertaining to
the use of hybrid density functionals.

\section*{References}
\bibliographystyle{unsrt}
\bibliography{refs}

\begin{thebibliography}{10}

\bibitem{Seshadri:2001}
R.~Seshadri.
\newblock {\em Proc. Indian Acad. Sci.}, 113:487--496, 2001.

\bibitem{Hegde2009}
M.~S. Hegde, Giridhar Madras, and K.~C. Patil.
\newblock {\em Acc. Chem. Res.}, {42}:{704--712}, {2009}.

\bibitem{Keezer:1968}
R.~C. Keezer, D.~L. Bowman, and J.~H. Becker.
\newblock {\em J. Appl. Phys.}, 39:2062--2066, 1968.

\bibitem{Moore:1941}
W.~J. Moore and L.~Pauling.
\newblock {\em J. Am. Chem. Soc.}, 63:1392--1394, 1941.

\bibitem{Rogers:1971}
D.~B. Rogers, R.~D. Shannon, and J.~L. Gillson.
\newblock {\em J. Solid State Chem.}, 3:314--316, 1971.

\bibitem{Hulliger:1965}
F.~Hulliger.
\newblock {\em J. Phys. Chem. Solids}, 26:639--645, 1965.

\bibitem{Okamoto:1967}
H.~Okamoto and T.~Aso.
\newblock {\em Jap. J. Appl. Phys.}, 6:779, 1967.

\bibitem{Pillo:1997}
T~Pillo, R~Zimmermann, P~Steiner, and S~Hufner.
\newblock {\em J. Phys. Condens. Matter}, {9}:{3987--3999}, {1997}.

\bibitem{Nilsson:1979}
P.~O. Nilsson and M.~S. Shivaraman.
\newblock {\em J. Phys. C. Solid State Phys.}, 12:1423--1427, 1979.

\bibitem{REY:1978}
E.~Rey, M.~R. Kamal, R.~B. Miles, and B.~S.~H. Royce.
\newblock {\em J. Mater. Sci.}, {13}:{812--816}, {1978}.

\bibitem{Belsky:2002}
A.~Belsky, M.~Hellenbrandt, V.~L. Karen, and P.~Luksch.
\newblock {\em Acta Cryst. B}, 58:364--369, 2002.

\bibitem{MEYER:1978}
H.~Meyer and H.~M\"uller-Buschbaum.
\newblock {\em Z. Anorg. Allg. Chem.}, {442}:{26--30}, {1978}.

\bibitem{Kurzman:2010fk}
J.~A. Kurzman, X.~Ouyang, W.~B. Im, J.~Li, J.~Hu, S.~L. Scott, and R.~Seshadri.
\newblock {\em Inorg. Chem.}, 49:4670--4680, 2010.

\bibitem{PARK:1994}
K.~T. Park, D.~L. Novikov, V.~A. Gubanov, and A.~J. Freeman.
\newblock {\em Phys. Rev. B}, {49}:{4425--4431}, {1994}.

\bibitem{HASS:1992}
K.~C. Hass and A.~E. Carlsson.
\newblock {\em {Phys. Rev. B}}, {46}:{4246--4249}, {1992}.

\bibitem{AHUJA:1994}
R.~Ahuja, S.~Auluck, B.~Johansson, and M.~A. Khan.
\newblock {\em {Phys. Rev. B}}, {50}:{2128--2132}, {1994}.

\bibitem{Uddin:2005}
J.~Uddin, J.~E. Peralta, and G.~E. Scuseria.
\newblock {\em Phys. Rev. B}, 71:155112, 2005.

\bibitem{Heyd:2003}
J.~Heyd, G.~E. Scuseria, and M.~Ernzerhof.
\newblock {\em J. Chem. Phys.}, 118:8207, 2003.

\bibitem{Seriani:2009}
N.~Seriani, J.~Harl, F.~Mittendorfer, and G.~Kresse.
\newblock {\em J. Chem. Phys.}, 131:054701, 2009.

\bibitem{Bruska:2011}
M.~K. Bruska, I.~Czekaj, B.~Delley, J.~Mantzaras, and A.~Wokaun.
\newblock {\em Phys. Chem. Chem. Phys.}, 13:15947--15954, 2011.

\bibitem{Wang:2008}
X.~L. Wang.
\newblock {\em Phys. Rev. Lett.}, {100}:{156404}, {2008}.

\bibitem{Chen:2011}
S.~W. Chen, S.~C. Huang, G.~Y. Guo, J.~M. Lee, S.~Chiang, W.~C. Chen, Y.~C.
  Liang, K.~T. Lu, and J.~M. Chen.
\newblock {\em Appl. Phys. Lett.}, {99}:012103, {2011}.

\bibitem{Bennett:2010}
J.~W. Bennett, I.~Grinberg, P.~K. Davies, and A.~M. Rappe.
\newblock {\em Phys. Rev. B}, 82:184106, 2010.

\bibitem{Ozawa:2005}
T.~C. Ozawa, T.~Taniguchi, Y.~Nagata, Y.~Noro, T.~Naka, and A.~Matsushita.
\newblock {\em J. Alloys Compd.}, {388}:{1--5}, {2005}.

\bibitem{URIU:1991}
R.~Uriu, D.~Shimada, and N.~Tsuda.
\newblock {\em J. Phys. Soc. Jpn.}, {60}:{2479--2480}, {1991}.

\bibitem{Ozawa:2005a}
T.~C. Ozawa, T.~Taniguchi, Y.~Nagata, Y.~Noro, T.~Naka, and A.~Matsushita.
\newblock {\em J. Alloys Compd.}, {395}:{32--35}, {2005}.

\bibitem{Lee:2010}
K.~J. Lee, S.~M. Choo, J.~B. Yoon, K.~M. Song, Y.~Saiga, C.~Y. You, N.~Hur,
  S.~I. Lee, T.~Takabatake, and M.~H. Jung.
\newblock {\em J. Appl. Phys.}, {107}:{09C306}, {2010}.

\bibitem{Lee:2011}
K.~J. Lee, S.~M. Choo, Y.~Saiga, T.~Takabatake, and M.-H. Jung.
\newblock {\em J. Appl. Phys.}, {109}:{07C316}, {2011}.

\bibitem{Wang:2009}
X.~Wang, G.~Peleckis, C.~Zhang, H.~Kimura, and S.~Dou.
\newblock {\em Adv. Mater.}, {21}:2196--2199, {2009}.

\bibitem{Su:2011}
H.~L. Su, Huang.~S. Y., Y.~F. Chiang, J.~C.~A. Huang, C.~C. Kuo, Y.~W. Du,
  Y.~C. Wu, and R.~Z. Zuo.
\newblock {\em Appl. Phys. Lett.}, 99:102508, 2011.

\bibitem{Heyd:2004}
J.~Heyd and G.~E. Scuseria.
\newblock {\em J. Chem. Phys.}, 121:1187--1192, 2004.

\bibitem{Heyd:2005}
J.~Heyd, J.~E. Peralta, G.~E. Scuseria, and R.~L. Martin.
\newblock {\em J. Chem. Phys.}, 123:174101, 2005.

\bibitem{Janesko:2009}
B.~G. Janesko, T.~M. Henderson, and G.~E. Scuseria.
\newblock {\em Phys. Chem. Chem. Phys.}, 11:443--454, 2009.

\bibitem{Henderson:2011}
T.~M. Henderson, J.~Paier, and G.~E. Scuseria.
\newblock {\em Phys. Status Solidi B}, 248:767--774, 2011.

\bibitem{Kresse:1994fk}
G.~Kresse and J.~Hafner.
\newblock {\em Phys. Rev. B}, 49:14251, 1994.

\bibitem{Kresse:1996uq}
G.~Kresse and J.~Furthm\"uller.
\newblock {\em Phys. Rev. B}, 54:11169, 1996.

\bibitem{Blochl:1994vn}
P.~E. Bl\"ochl.
\newblock {\em Phys. Rev. B}, 50:17953, 1994.

\bibitem{Kresse:1999}
G~Kresse and D~Joubert.
\newblock {\em Phys. Rev. B}, {59}:{1758--1775}, {1999}.

\bibitem{Perdew:1996kx}
J.~P. Perdew, K.~Burke, and M.~Ernzerhof.
\newblock {\em Phys. Rev. Lett.}, 77:3865, 1996.

\bibitem{Krakau:2006}
A.~V. Krakau, O.~A. Vydrov, A.~F. Izmaylov, and G.~E. Scuseria.
\newblock {\em J. Chem. Phys.}, 125:224106, 2006.

\bibitem{Andersen:1975}
O.~K. Andersen.
\newblock {\em Phys. Rev. B}, 12:3060, 1975.

\bibitem{Jepsen:1995}
O.~Jepsen and O.~K. Andersen.
\newblock {\em Z. Phys. B}, 97:35, 1995.

\bibitem{Jepsen:2000}
O.~Jepsen and O.~K. Andersen.
\newblock {The Stuttgart TB-LMTO-ASA Program version 47, MPI f\"ur
  Festk\"orperforschung, Stuttgart, Germany}.
\newblock 2000.

\bibitem{Barth:1972}
U.~von Barth and L.~Hedin.
\newblock {\em J. Phys. C}, 5:1629, 1972.

\bibitem{Becke:1990}
A.~D Becke and K.~E. Edgecombe.
\newblock {\em J. Chem. Phys.}, 92:5397, 1990.

\bibitem{Silvi:1994}
B.~Silvi and A.~Savin.
\newblock {\em Nature}, 371:683, 1994.

\bibitem{Watson:1999}
G.~W. Watson, S.~C. Parker, and G.~Kresse.
\newblock {\em Phys. Rev. B}, 59:8481--8486, 1999.

\bibitem{Watson:1999a}
G.~W. Watson and S.~C. Parker.
\newblock {\em J. Phys. Chem. B}, 103:1258--1262, 1999.

\bibitem{Raulot:2002}
J.-M. Raulot, G.~Baldinozzi, R.~Seshadri, and P.~Cortona.
\newblock {\em Solid State Sci.}, 4:467--474, 2002.

\end{thebibliography}

\end{document}